\documentclass[sigconf]{acmart}

\usepackage{booktabs} 

\usepackage[bottom]{footmisc}

\copyrightyear{2018}
\acmYear{2018}
\setcopyright{acmcopyright}
\acmConference[GLSVLSI '18]{2018 Great Lakes Symposium on VLSI}{May
23--25, 2018}{Chicago, IL, USA}
\acmBooktitle{GLSVLSI '18: 2018 Great Lakes Symposium on VLSI, May
23--25, 2018, Chicago, IL, USA}
\acmPrice{15.00}
\acmDOI{10.1145/3194554.3194625}
\acmISBN{978-1-4503-5724-1/18/05}

\vspace{-10.5in}

\usepackage{kantlipsum}
\makeatletter
\def\@maketitle{\newpage
 \null
 \setbox\@acmtitlebox\vbox{%
\baselineskip 20pt
\vskip -2cm                  
   \begin{center}
    {\ttlfnt \@title\par}       
    \vskip -2em                
{\subttlfnt \the\subtitletext\par}\vskip 1.25em
    {\baselineskip 18pt\aufnt   
     \lineskip 0em             
     \begin{tabular}[t]{c}\@author
     \end{tabular}\par}
    \vskip 0em               
   \end{center}}
 \dimen0=\ht\@acmtitlebox
 \unvbox\@acmtitlebox
 \ifdim\dimen0<0.0pt\relax\vskip-\dimen0\fi}

\begin{document}
\title{Structured Weight Matrices-Based Hardware Accelerators in Deep Neural Networks: FPGAs and ASICs}



\vspace{-0.3in}
\author{Caiwen Ding\footnotemark[1]$^1$, Ao Ren\footnotemark[1]$^1$, Geng Yuan$^1$, Xiaolong Ma$^1$, Jiayu Li$^1$, Ning Liu$^1$, Bo Yuan$^2$, Yanzhi Wang$^1$}
\thanks{$^{*}$Caiwen Ding and Ao Ren contributed equally to this work.}
\affiliation{%
  \institution{
  $^1$Syracuse University, $^2$City University of New York, City College.}
}
\email{{cading, aren,  geyuan, xma27, jli221, nliu03, ywang393}@syr.edu, byuan@ccny.cuny.edu}

\renewcommand{\shortauthors}{C. Ding et al.}

\begin{abstract}
Both industry and academia have extensively investigated hardware accelerations. In this work, to address the increasing demands in  computational capability and memory requirement,
 we propose structured weight matrices (SWM)-based compression techniques for both \emph{field programmable gate array} (FPGA) and \emph{application-specific integrated circuit} (ASIC) implementations. In algorithm part, SWM-based framework adopts block-circulant matrices to achieve a fine-grained tradeoff between accuracy and compression ratio. 
The SWM-based technique can reduce computational complexity from O($n^2$) to O($n\log n$) and storage complexity from O($n^2$) to O($n$) for each layer and both training and inference phases. 
For FPGA implementations on deep convolutional neural networks (DCNNs), we achieve at least 152X and 72X improvement in performance and energy efficiency, respectively using the SWM-based framework, compared with the baseline of IBM TrueNorth processor under same accuracy constraints using the data set of MNIST, SVHN, and CIFAR-10. For FPGA implementations on long short term memory (LSTM) networks, the proposed SWM-based LSTM can achieve up to 21X enhancement in performance and 33.5X gains in energy efficiency compared with the baseline accelerator. For ASIC implementations,  the SWM-based ASIC design exhibits impressive advantages in terms of power, throughput, and energy efficiency.
Experimental results indicate that this method is greatly suitable for applying DNNs onto both FPGAs and mobile/IoT devices.
\end{abstract}

%
%

\ccsdesc[500]{Computer systems organization~Embedded systems}

\keywords{Deep learning, FPGA, ASIC, Accelerator, Structured weight matrices}

\maketitle

\section{Introduction}

Deep learning has  increasingly drawn attentions in many research fields, such as speech recognition~\cite{hinton2012deep}, computer vision~\cite{krizhevsky2012imagenet,he2016deep}, self-driving cars~\cite{schmidhuber2015deep,huval2015empirical}, and unmanned aircraft systems~\cite{makantasis2015deep}. Large-scale deep neural networks (DNNs) typically consist of multiple layers, and at least millions of weight parameters for the entire model \cite{krizhevsky2012imagenet}. One major advantage of the larger-scale DNNs is that they extract more complex high-level features from the inputs (e.g., images/videos, speeches), and as a result, achieving a significant improvement in model accuracy \cite{schmidhuber2015deep}. 

On the other hand, as the size of DNNs grows continuously, there exist tremendous demands in increasing computational capability and memory requirement. Therefore, improving the performance and energy efficiency while maintaining the accuracy of DNNs becomes extremely critical. Two trends have characterized the research advance in order to achieve higher performance and energy efficiency. The first trend is {\em hardware acceleration}. FPGA-based accelerators have the advantage of friendly programmability and high-degree parallelism. Stochastic Computing (SC), in which all the inputs and weight values are represented as streams of random bits, has been investigated and successfully applied to hardware acceleration of DNNs~\cite{li2016dscnn,li2017towards,ren2017sc,Ao:DCNN1,yuan2017softmax,li2017hardware,li2017normalization,lin2017fft,li2017structural}. Data-path optimization technique~\cite{gokhale2014240} have also been studied to map a limited number of Processing elements (PEs) on FPGA and reuse the mapped PEs by iterating data through them. On the other hand, ASIC-based implementations have been explored to further accelerate DNNs. A substantial number of high-tech companies  have declared their ASIC chip designs in DNNs such as Google~\cite{jouppi2017datacenter}
and IBM TrueNorth~\cite{esser2015backpropagation}. In the field of academia, Eyeriss \cite{chen2017eyeriss}, EIE \cite{han2016eie}, and DaDianNao~\cite{chen2014dadiannao} mainly focus on the convolutional layers, the fully-connected layers, and the memory design/organization at the architectural level, respectively.

The second trend is {\em model compression} motivated by energy efficiency limitation of large DNN models. Weight pruning~\cite{han2015deep} and lower rank approximation~\cite{tai2015convolutional} have aimed to the reduce the number of operations involved in
DNNs. They achieve a parameter reduction to some extent with inconsequential accuracy degradation. However, they have brought the new challenges into DNNs such as irregular network structure caused by sparsity regularization \cite{yu2017scalpel}, and increased training complexity caused by the
additional pruning process \cite{han2015deep} or low rank approximation step \cite{tai2015convolutional}.

In this work, to address the limitations of existing works in model size compression and acceleration and to achieve ultra-high energy efficiency and performance for FPGA and ASIC-based hardware implementations, we propose the structured weight matrices (SWM)-based compression technique on both FPGA and ASIC implementations. The SWM-based framework adopts the general block-circulant matrices to achieve a fine-grained tradeoff between accuracy and compression ratio.
For FPGA implementations on DCNNs, we achieve at least 152X and 72X improvement in performance and energy efficiency, respectively using SWM-based framework, compared with the baseline of IBM TrueNorth processor under same accuracy constraints using the data set of MNIST, SVHN, and CIFAR-10. For FPGA implementations on LSTM networks, the proposed SWM-based method can achieve up to 21X enhancement in performance and 33.5X gains in energy efficiency compared with ESE, respectively. For ASIC implementations,  the proposed SWM-based design exhibits impressive advantages in terms of power, throughput, and energy efficiency. It indicates that this method is greatly suitable for applying DNNs onto both FPGAs and mobile/IoT devices.

\section{Background of DNNs}


\subsection{Deep Convolutional Neural Networks}
DNN systems consist of many different architectures such as DCNNs, recurrent neural network (RNNs), and deep belief networks (DBNs). Although different network structures target at specific applications, they have the similarity in construction principle, i.e., multiple layers connected in series for feature extraction~\cite{lee2009convolutional,karpathy2014large}. DNNs are commonly made up of three-layer types: Fully-connected (FC) and convolutional layers (CONV), and pooling layers (POOL). 

\emph{\textbf{FC layer}} 
is the most storage-intensive layer in DNNs \cite{qiu2016going,han2016eie} since its neurons are fully connected with neurons in previous layer. The computation of an FC layer consists of matrix-vector arithmetics followed by the activation function, described as:
$\bf{y}=\psi(\bf{W}\bf{x}+\mathbf{\theta})$,
where $\mathbf{W}\in \mathbb{R}^{m\times n}$ is the weight matrix of the synapses between this FC layer (with $m$ neurons) and its previous layer (with $n$ neurons); $\mathbf{\theta}\in \mathbb{R}^{m}$ is the bias vector; and $\psi(\cdot)$ is the activation function.
The calculation of $\bf{Wx}$ dominates computational complexity because the rest has lower complexity of O($n$).

\emph{\textbf{CONV layer}} performs a multi-dimensional convolution to extract features from its inputs that will be fed into subsequent layers for extracting higher-level features. A CONV layer is associated with a set of learnable filters (or kernels) \cite{lecun1998gradient}. A filter-sized moving window is applied to the input feature maps, calculating the convolution of the filter and input feature maps in the moving window. In practical DNN models, the CONV layers are often associated with multiple input and multiple output feature maps. As a result, the CONV layer can be expressed in \emph{tensor computations}:
$\mathcal{Y}(x,y,p)=\sum_{i=1}^r\sum_{j=1}^r\sum_{c=1}^C\mathcal{F}(i,j,c,p)\mathcal{X}(x+i-1,y+j-1,c),$
where $\mathcal{X}\in\mathbb{R}^{W\times H\times C}$, $\mathcal{Y}\in\mathbb{R}^{(W-r+1)\times(H-r+1)\times P}$, $\mathcal{F}\in\mathbb{R}^{r\times r\times C\times P}$ represent the input, output, and weight ``tensors" of the CONV layer, respectively. Here, $W$ and $H$ are the spatial dimensions of the input maps, $C$ is the number of input maps, $r$ is the size of the convolutional kernel, and $P$ is the number of output maps.
 
 \emph{\textbf{POOL layer}} performs a subsampling operation on the extracted features to reduce the data dimensions and mitigate overfitting issues. Max pooling is the dominant type of pooling strategy in state-of-the-art DCNNs due to its higher overall accuracy and convergence speed \cite{chen2014dadiannao,chen2017eyeriss}.
 
The majority of computations occur in CONV and FC layers, while the POOL layer has a lower computational complexity of O($n$). The storage requirement of DNNs is due to the weight matrices $\textbf{W}$'s in the FC layers and the convolutional kernels $\textbf{F}$'s in CONV layers. As a result, the FC and CONV layers become the major research focuses for energy-efficient implementation of DNNs.

\subsection{Recurrent Neural Networks}

 \begin{figure} [b]
  \centering
  \includegraphics[width=0.8\columnwidth]{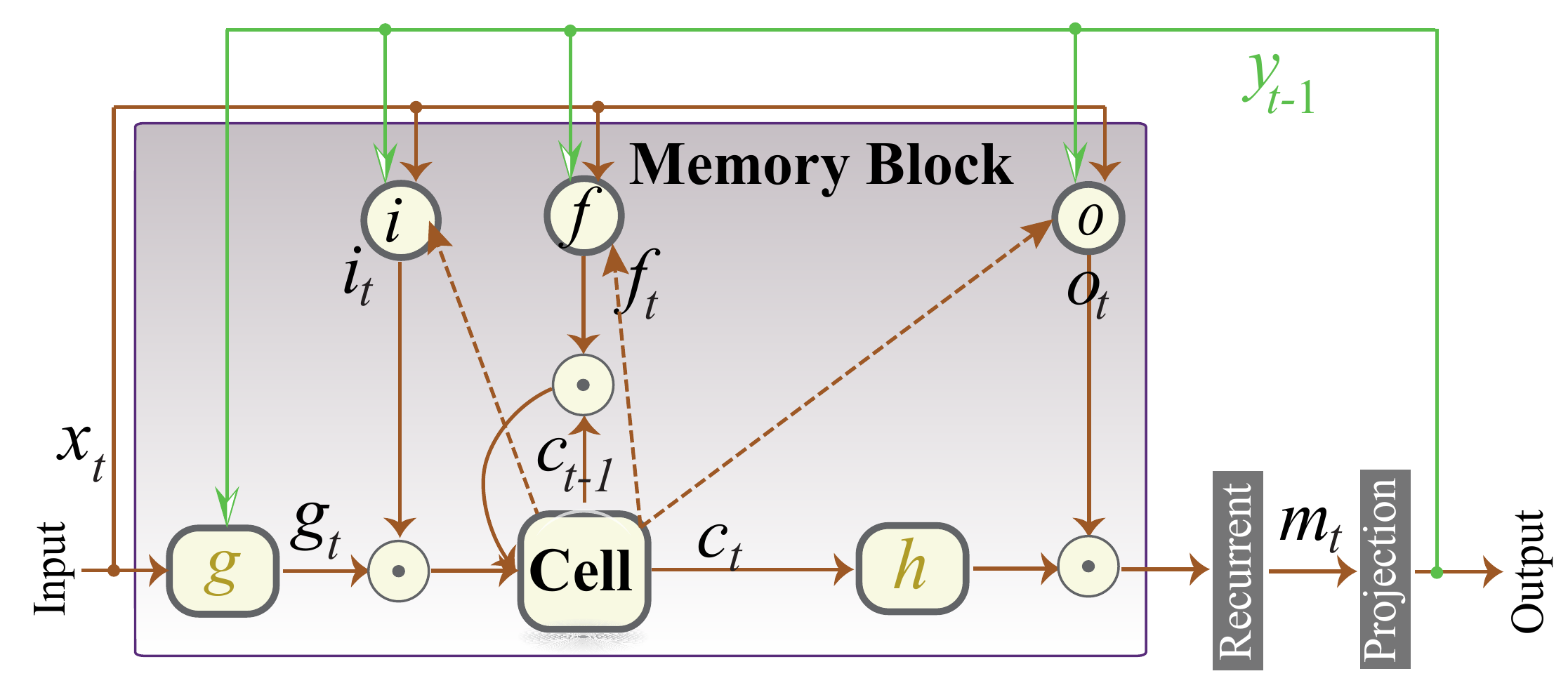}  
  \caption{An example of LSTM based RNN architecture.}
  \label{fig:LSTM}
 \end{figure}

RNNs have been investigated and have many applications in natural language processing, speech recognition, and machine translation~\cite{sak2014long}. As one popular type of RNNs, long short term memory (LSTM) has been broadly studied as shown in Fig.~\ref{fig:LSTM}~\cite{sak2014long}. An LSTM-based RNN accepts an input sequence $\mathbb{X}= (\mathbf{x}_1; \mathbf{x}_2; \mathbf{x}_3; ...; \mathbf{x}_T)$ (each of $\mathbf{x}_t$ is a vector corresponding to time $t$) with the output sequence from last step $\mathbb{Y}^{T-1} = (\mathbf{y}_0; \mathbf{y}_1; \mathbf{y}_2; ...; \mathbf{y}_{T-1})$ (each of $\mathbf{y}_t$ is a vector).
It computes an output sequence $\mathbb{Y} = (\mathbf{y}_1;\mathbf{y}_2; \mathbf{y}_3; ...; \mathbf{y}_T )$ by using the following equations iteratively from $t = 1$ to $T$:
\begin{subequations}\label{eqn:model}
\begin{align}
    \mathbf{i}_t &= \sigma(\mathbf{W}_{ix}\mathbf{x}_t +\mathbf{W}_{ir}\mathbf{y}_{t-1} + \mathbf{W}_{ic}\mathbf{c}_{t-1}+\mathbf{b}_i), \\
    \mathbf{f}_t &= \sigma(\mathbf{W}_{fx}\mathbf{x}_t +\mathbf{W}_{fr}\mathbf{y}_{t-1} + \mathbf{W}_{fc}\mathbf{c}_{t-1}+\mathbf{b}_f), \\
    \mathbf{g}_t &= \sigma(\mathbf{W}_{cx}\mathbf{x}_t + \mathbf{W}_{cr}\mathbf{y}_{t-1} + \mathbf{b}_c), \\
    \mathbf{c}_t &= \mathbf{f}_t \odot \mathbf{c}_{t-1} + \mathbf{g}_t \odot \mathbf{i}_t, \\
    \mathbf{o}_t &= \sigma(\mathbf{W}_{ox}\mathbf{x}_t + \mathbf{W}_{or}\mathbf{y}_{t-1} + \mathbf{W}_{oc}\mathbf{c}_{t}+\mathbf{b}_o), \\
    \mathbf{m}_t &= \mathbf{o}_t \odot \mathbf h(\mathbf{c}_t), \\
    \mathbf{y}_t &= \mathbf{W}_{ym}\mathbf{m}_t,
\end{align}
\end{subequations} 

where symbols $\mathbf{i}$, $\mathbf{f}$, $\mathbf{o}$, $\mathbf{c}$, $\mathbf{m}$, and $\mathbf{y}$ represent the input gate, forget gate, output gate, cell state, cell output, and projected output, respectively. The $\odot$ operation represents element-wise multiplication, and the $+$ operation is matrix addition.
The $\mathbf{W}$ terms represent weight matrices (for instance, $\mathbf{W}_{ix}$ is the weight matrix from the input vector $\mathbf{x}_t$ to the input gate), and the $\mathbf{b}$ terms are the bias vectors. 
Additionally, weight matrices $\mathbf{W}_{ic}$, $\mathbf{W}_{fc}$, and $\mathbf{W}_{oc}$ are diagonal matrices for peephole connections, which can be considered as vectors during matrix-vector multiplication. Therefore, $\mathbf{W}_{ic}\mathbf{c}_{t-1}$ can be calculated using $\odot$ operation.
$\sigma$ is the logistic activation function and $h$ is a self-defined activation function. In this model we use hyperpolic tangent (tanh) activation function as $h$.


\section{Structured Weight Matrix}
\label{sec:design meth}
\begin{figure*}[!t]
\centering
\includegraphics[width = 1.6\columnwidth]{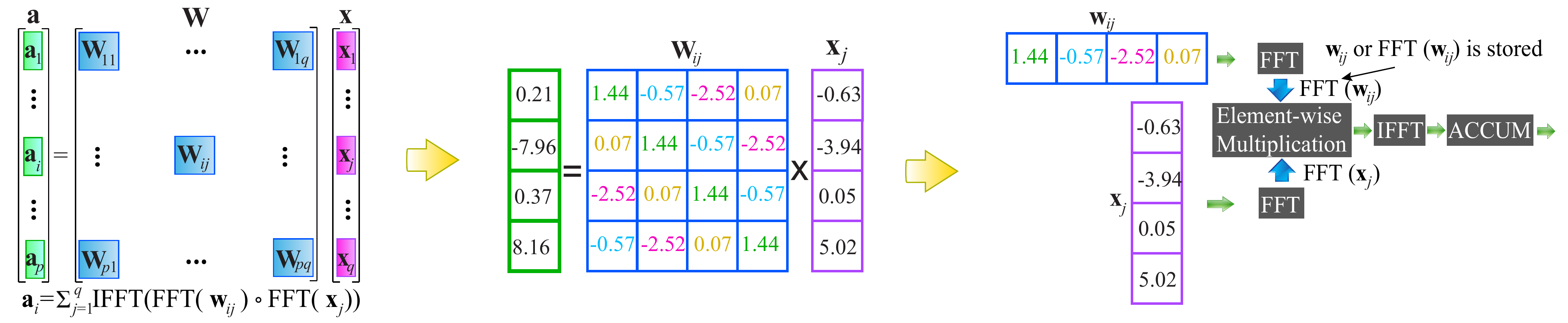}
\vskip -0.8em
\caption{FFT-Based Calculation in SWM-based FC Layer.}
\label{fig:Block}
\end{figure*}

This section discusses the inference and training algorithms of SWM-based DNNs
(e.g.,~\cite{ding2017c,wang2018towards}). 
The advantage is two-fold:
{\em 1)} it is possible to derive a fine-grained tradeoff between accuracy 
and compression/acceleration by changing the block size; and 
{\em 2)} the method applies to both FC and CONV layers. 
The theoretical foundation is also derived from \cite{zhao2017icml}, which shows that the ``effectiveness'' of SWM-based DNNs is the same compared with DNNs without compression. Experimental results in \cite{ding2017c,wang2018towards} have demonstrated a good ratio of model compression (i.e., from 41$\times$ to 256$\times$) with small (less than 2\%) overall accuracy degradation.
In the following, we discuss the inference and training algorithms for FC layer,
details of the CONV layer algorithms are provided in \cite{ding2017c}.

The key idea of SWM-based FC layers is to partition the original weight matrix $\textbf{W}\in \mathbb{R}^{m\times n}$ into blocks of square sub-matrices, and each sub-matrix is a circulant matrix. 
The illustrations are shown in Fig. \ref{fig:Block}. 
Let $k$ denote the \emph{block size} (size of each sub-matrix) and 
assume there are $p \times q$ blocks after partitioning $\mathbf{W}$, where $p = m\div k$ and $q=n \div k$. Then $\mathbf{W} = [\mathbf{W}_{ij}]$, $i \in \{1 \dots p\}$, $j \in \{1 \dots q\}$. The input $\mathbf{x}$ is also partitioned as $\mathbf{x} = [\mathbf{x}^T_1, \mathbf{x}^T_2, \dots, \mathbf{x}^T_q]^T$. Then, the \emph{forward propagation} of FC layer in the inference is given by (with bias and ReLU omitted for simplicity):
\begin{equation}
\mathbf{a}
=
\mathbf{Wx} 
=
\begin{bmatrix}
         \sum_{j=1}^q \mathbf{W}_{1j} \mathbf{x}_j   \\
         \sum_{j=1}^q \mathbf{W}_{2j} \mathbf{x}_j   \\
         \dots \\
         \sum_{j=1}^q \mathbf{W}_{pj} \mathbf{x}_j  
\end{bmatrix}
=
\begin{bmatrix}
         \mathbf{a}_1   \\
         \mathbf{a}_2   \\
         \dots \\
         \mathbf{a}_p
\end{bmatrix},
\end{equation}
where $\mathbf{a}_i \in \mathbb{R}^{k}$ is a column vector. Assume each circulant matrix $\mathbf{W}_{ij}$ is defined by a vector $\mathbf{w}_{ij}$, i.e., $\mathbf{w}_{ij}$ is the first row vector of $\mathbf{W}_{ij}$. According to the \emph{circulant convolution theorem} \cite{pan2012structured}, the calculation of $\mathbf{W}_{ij} \mathbf{x}_j$ can be performed as $\text{IFFT}\big(\text{FFT}(\mathbf{w}_{ij})\circ\text{FFT}(\mathbf{x}_j)\big)$, where $\circ$ denotes element-wise multiplications. The operation procedure is shown on the right of Fig. \ref{fig:Block}. For the inference phase, the computational complexity of this FC layer is $O(pqk\log k)$, which is equivalent to $O(n\log n)$ for small $p$, $q$ values. Similarly, the storage complexity is $O(pqk)$ because only $\mathbf{w}_{ij}$ or $\text{FFT}(\mathbf{w}_{ij})$ for each sub-matrix needs to be stored, which is equivalent to $O(n)$ for small $p$, $q$ values. 
Therefore, the simultaneous acceleration and model compression are achieved.

\section{MODEL COMPRESSION and Accuracy}

To reduce the computation complexity and storage complexity, many researchers have investigated to reduce the number of weight parameters or the number of bits for weight representation. However, the compression techniques will cause the model accuracy degradation. 
In this section, we will discuss the trade-off between model compression and model accuracy loss of the SWM-based technique.

\subsection{Quantization and Weight Reduction}

Data quantization on weights and neurons is a commonly used method for model compression. We attempt to use low-bit fixed-point data to represent the neurons and weights instead of using floating point data. We design a bit-wise simulator using C++ to verify the total number of bits for both integer and fractional part.
Structure weight matrix, as a low-rank representation, uses one or several block circulant matrices to replace the original weight matrix as discussed in Section.~\ref{sec:design meth}. Shown in Fig.~\ref{fig:Block}, by partitioning the original weight matrix $\textbf{W}\in \mathbb{R}^{m\times n}$ into $p\times q$ blocks of square sub-matrices, the total number of weights are reduced from $m \times n$ to $\frac{m}{k}\times \frac{n}{k}\times k =(m\times n)/k$, where each block is a $k \times k$ matrix. We further investigate the SWM-based DNN models including DCNNs and LSTMs regarding the compression ratio (block size) and model accuracy.

\subsection{Accuracy Evaluation}
\subsubsection{Accuracy Evaluation on DCNNs} The weight storage (model size) reduction, and the test accuracy on various image recognition datasets and DCNN models: MNIST (LeNet-5), CIFAR-10, SVHN, STL-10, and ImageNet (using AlexNet structure) \cite{krizhevsky2012imagenet,netzer2011reading,krizhevsky2009learning,coates2010analysis,deng2012mnist}) are discussed in~\cite{ding2017c}. 16-bit data quantization is adopted and the baselines are the original DCNN models with unstructured weight matrices and 32-bit floating point representations. The 
SWM-based compression technique enables 400$\times$-4000+$\times$ reduction in model size in the corresponding FC layers. On the other hand, the accuracy is close to original DCNN models and the accuracy degradation is negligible.
Moreover, another advantage of the SWM-based technique is that
the storage process of weight parameter after compression is regular, while reference works \cite{han2015deep} bring in irregularity in storing the weight parameter. The introduced irregularity requires extra index per weight parameter and therefore affects the available parallelism degree.

\subsubsection{Accuracy Evaluation on LSTM}

We evaluate the structure matrices based compression technique using TIMIT benchmark, the most commonly used dataset for automatic speech recognition (ASR) application. The LSTM network is built by stacking multiple LSTM layers. The Google LSTM model~\cite{sak2014long} with unstructured weight matrix is selected as the baseline model. We preprocess the TIMIT audio data using FFT-based filterbank as discussed in~\cite{wang2018clstm,li2018efficient}. The input speech data have the same number of features and same architecture as ESE ~\cite{han2017ese}. Phone Error Rate (PER) is adopted to evaluate the model prediction accuracy.

The block-circulant matrix based LSTM  model enables a comprehensive tuning of model compression ratio by varying the block size $k$. The PER is close to baseline LSTM when the block size is 2 using SWM-based compression technique. For the SWM-based LSTM models with a block size of 8 and 16, 7.6X and 14.6X model size reduction can be achieved compared with baseline LSTM, respectively. On the other hand, the computational complexity is reduced by 2.6X and 3.7X while the PERs are only $0.32\%$ and $1.23\%$ higher than the baseline. 

\section{SWM-Based HARDWARE DESIGN}
\subsection{FPGA}

\subsubsection{Overall Architecture}
\begin{figure}[t]
\centering
\includegraphics[width = 0.7\columnwidth]{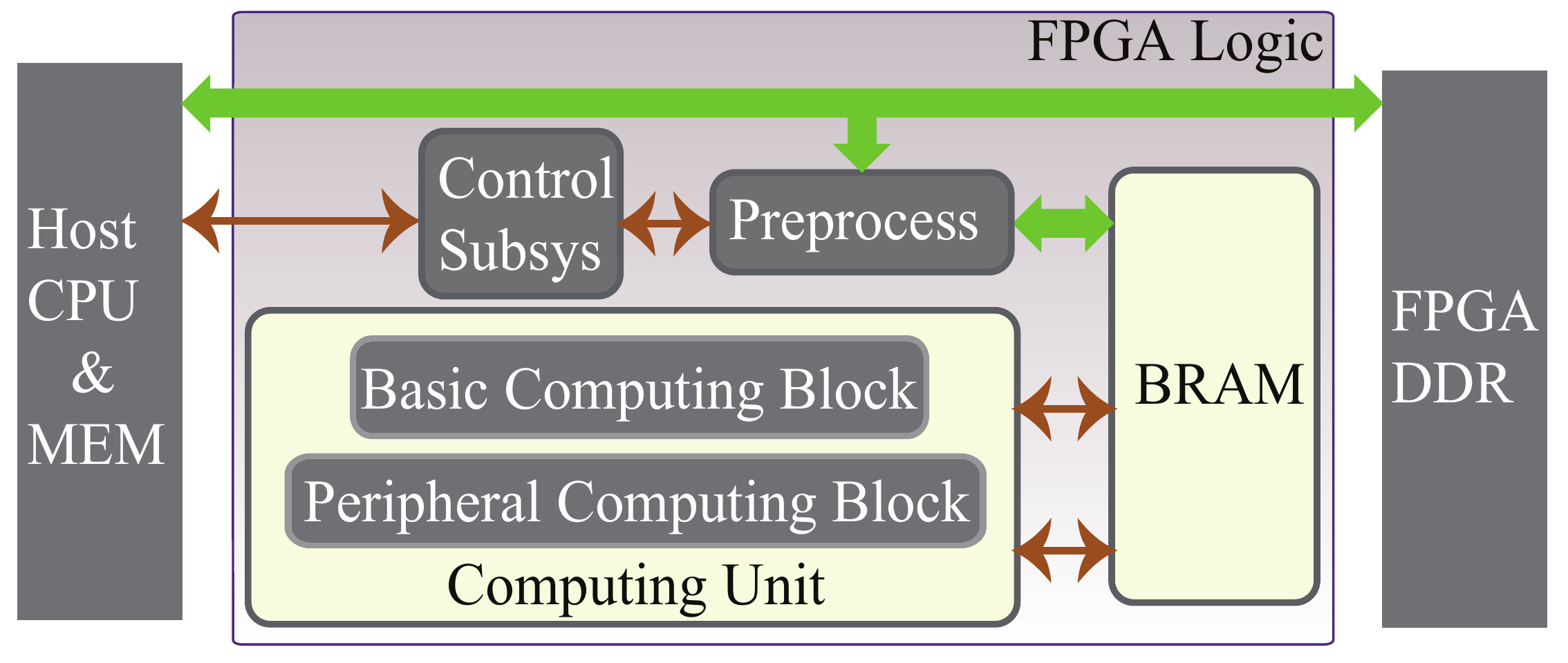}
\vspace{-0.1in}
\caption{Overall system architecture of the proposed SWM-based FPGA compression framework.}
\label{fig:overall_diagram}
\end{figure}
The overall SWM-based architecture is shown in Fig. \ref{fig:overall_diagram}. The Host CPU is responsible for issuing workload or instructions to the FPGA logic block and monitoring the working stats. The FPGA logic part includes computing unit (containing the basic computing block and the peripheral computing block), the control subsystem, BRAM block, and the preprocess block for certain designs when the data loaded from external memory requires preprocess. The memory hierarchy of the architecture primarily consists of three blocks: Host MEM, FPGA DDR, and on-chip block memory (BRAM).  The control subsystem coordinates the actual FFT/IFFT operations in the basic computing block and peripheral computing block. The control subsystem also determines the input size of FFT/IFFT operations. The twiddle factors in FFT/IFFT operations are stored in BRAM (i.e., the $W_n^i$ values including both real and imaginary parts); the weights, e.g., the FFT results $\text{FFT}(\mathbf{w}_{ij})$ are also stored in BRAM.

\subsubsection{Computing Unit Designs}

\begin{figure}[t]
\centering
\includegraphics[width = 1\columnwidth]{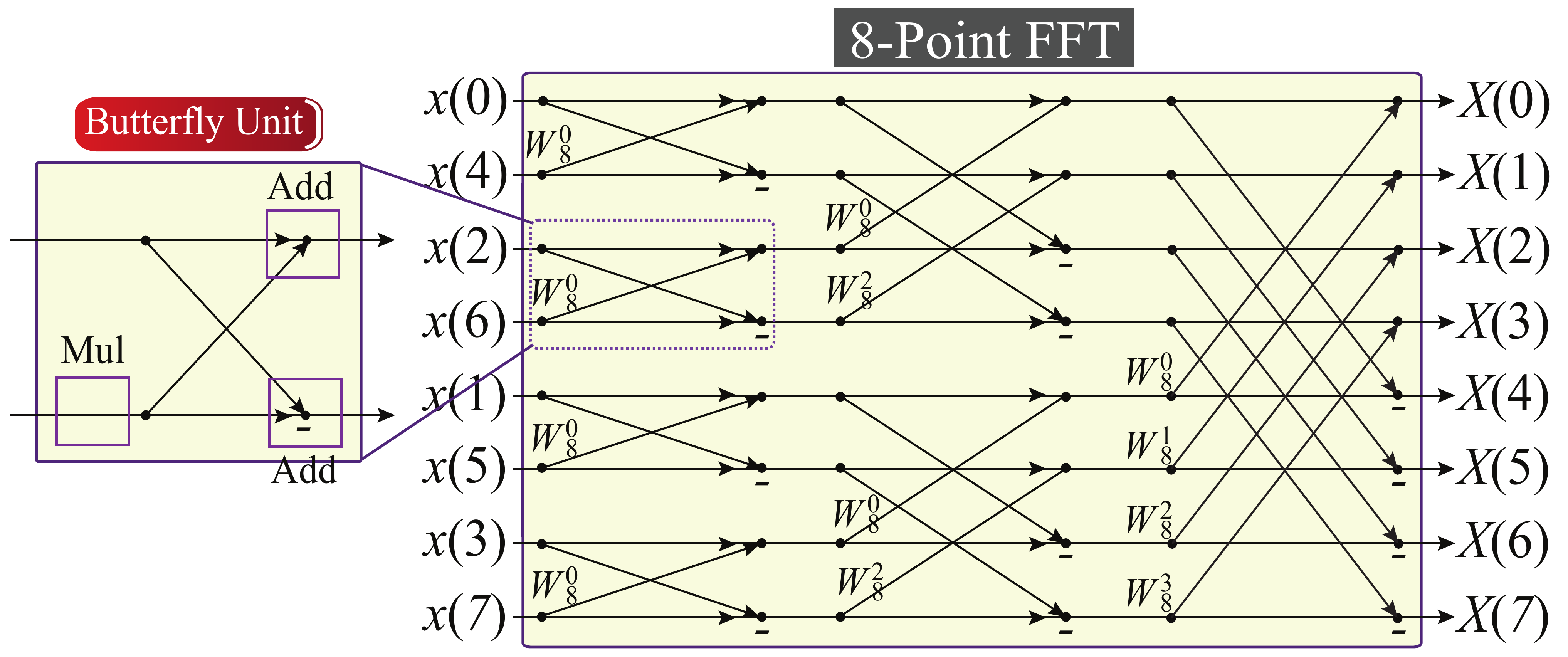}
\caption{An example of 8-point basic computing block for FFT using butterfly units.}
\label{fig:FFT}
\end{figure}

In the computing unit, the peripheral computing block mainly focuses on component-wise multiplication, activation (ReLU, Tanh, and Sigmoid), pooling etc., which need lower computational cost and hardware footprint.
The basic computing unit consists of an FFT operation with a parallelization degree of $N$ and depth of $\log N$. Fig.~\ref{fig:FFT} shows an example of 8-point FFT operation in the basic computing block using butterfly units. The IFFT operation can also be implemented using the $N$ inputs basic computing unit in addition to a division operation (i.e., $\div N$) and two conjugations.

\subsection{ASIC}
\begin{figure}[b]
    \centering
    \includegraphics[width=0.4\textwidth]{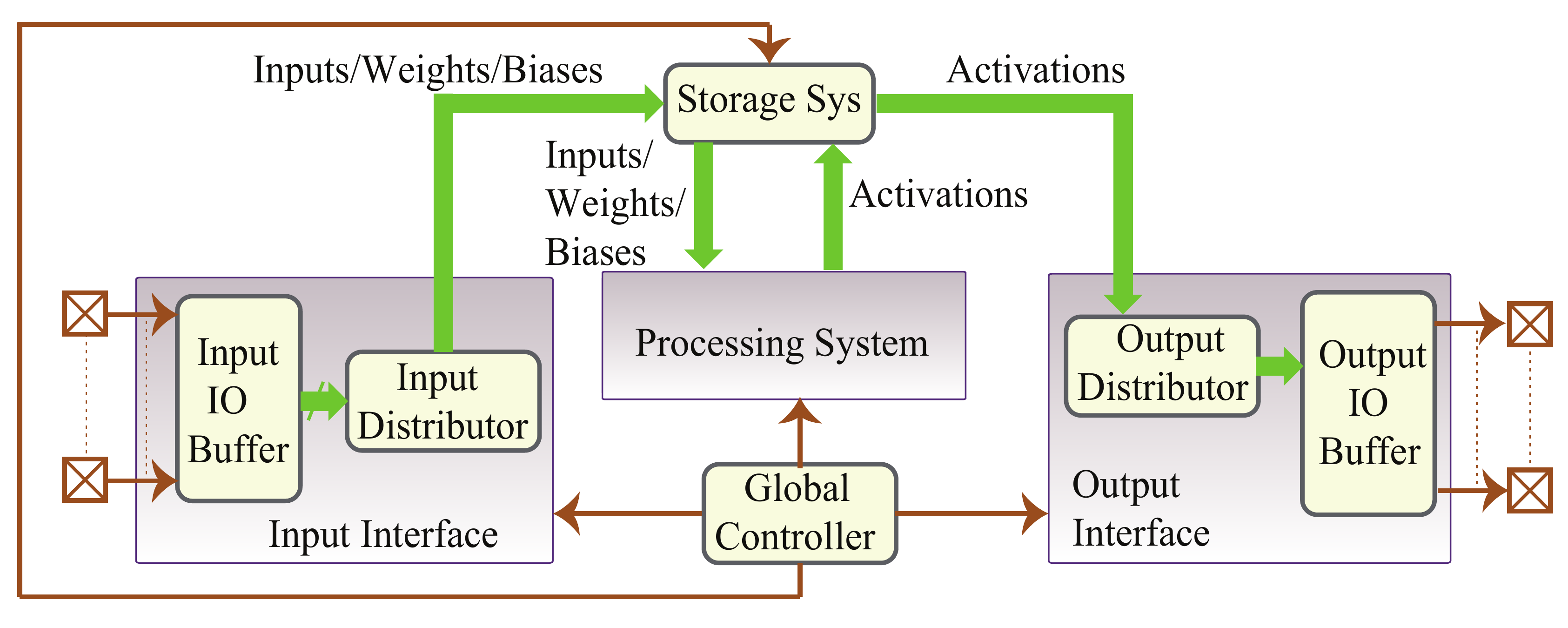}
    \caption{The architecture of the SWM-based chip.}
    \label{fig:asic_architecture}
\end{figure}

In order to apply DNNs onto mobile/IoT devices, the DNN applications should be implemented in ASICs, due to the benefit of small hardware volume. The great reduction in both parameter size and computational time complexity makes our SWM-based method suitable for ASIC implementations. Figure \ref{fig:asic_architecture} shows the architecture of our end-to-end ASIC implementation of the SWM-based DNNs. The architecture consists of four main blocks: input/output interface, storage system, processing system, and global controller. 

The input/output interface is in charge of communicating with the external environment of the chip and the on-chip storage system. The input interface is composed of an input IO buffer and an input distributor. Similarly, the output interface is composed of an output IO buffer and an output distributor. In the view of data flow, the input IO buffer first receives and buffers data, including input images, weights, and biases from the external environment. For the reason that the number of IO pads are usually limited to a small number, whereas the bandwidth of the processing system is rather large for achieving high parallelism of computation. This mismatch in bandwidth requires an input distributor to temporally hold the external data until the size of the data reaches the bandwidth requirement of the storage system. Besides, there are three storage modules inside the storage system for respectively storing inputs/intermediate activations, weights, and biases, the global controller will decide where the buffered data should flow. With the similar idea, the output distributor will receive final activations from the storage system and be controlled to distribute a portion of activations into the output IO buffer, which will further send them back to the external system.  

\begin{figure}[t]
    \centering
    \includegraphics[width=0.4\textwidth]{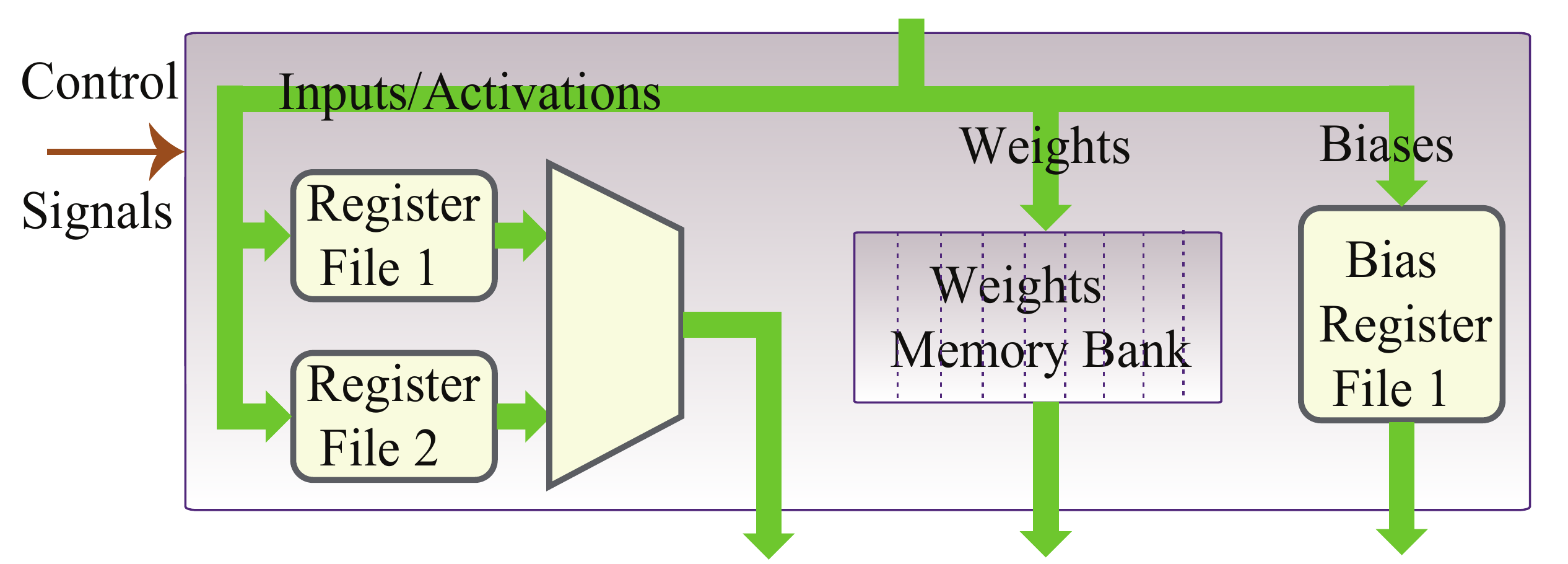}
    \caption{The architecture of the storage system.}
    \label{fig:storage_architecture}
\end{figure}

As depicted in Figure \ref{fig:storage_architecture}, the storage system composes three subsystems, including a memory bank for storing weights, a register file for storing biases, and a ping-pong buffer (i.e., two alternating register files) for storing image inputs and intermediate activations. 

\begin{figure}[b]
    \centering
    \includegraphics[width=0.5\textwidth]{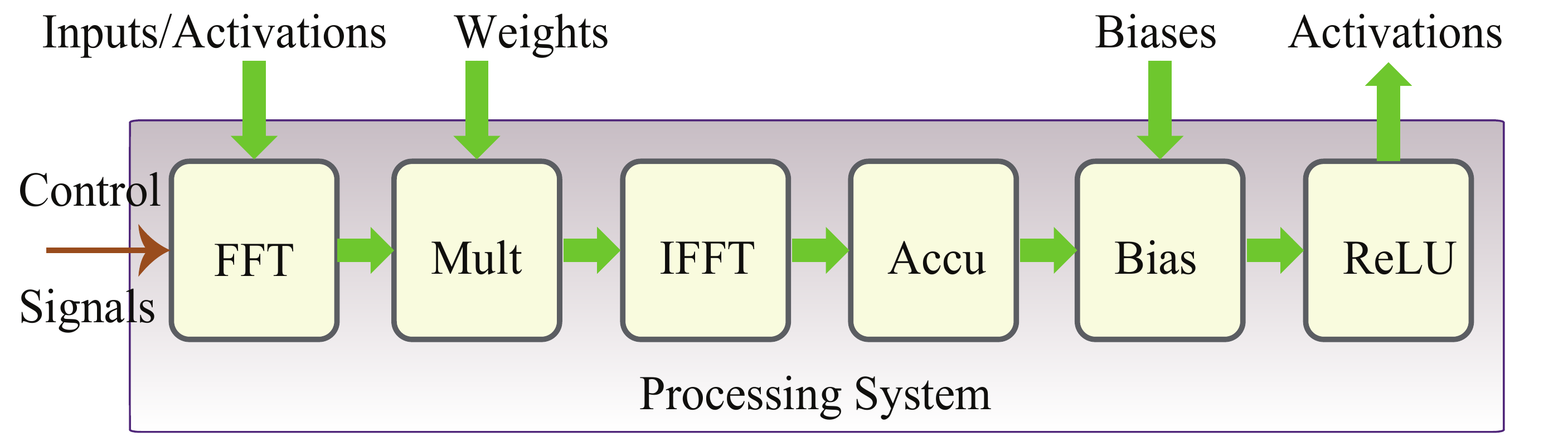}
    \caption{The architecture of the processing system.}
    \label{fig:process_architecture}
\end{figure}

The processing system achieves following equation for each layer: 
$\mathbf{y}_j = h(\Sigma\text{IFFT}\big(\text{FFT}(\mathbf{w}_{ij})\circ\text{FFT}(\mathbf{x}_j)\big) + \mathbf{b}_j)$, where $\mathbf{w}_{ij}$ is the vector of weights at the ith row and jth column of the weight matrix, $\mathbf{x}_j$ and $\mathbf{b}_j$ are respectively the jth vector of inputs/activations and biases, and $h(\cdot)$ is an activation function. According to above equation, the processing system should contain the modules that are illustrated in Fig. \ref{fig:process_architecture}. As the first step in the core computation, the image inputs are loaded from the storage system to the FFT module. Since the weights are repeatedly used without changes, what the weight memory bank stores are the weights in frequency domain. Thus the inputs of the multiply module are $\text{FFT}(\mathbf{x}_j)$ and $\text{FFT}(\mathbf{w}_{ij})$. Next, the $\text{IFFT}$ module performs the inverse $\text{FFT}$ operation over the element-wise production vector, converting the vector from frequency domain to time domain. Then the summation is performed by the Accumulator module that generates the dot-product of inputs and weights. Finally, the Biase module adds up the biases to the dot-products, and the Activation module produces a vector of activations.

Another crucial module in the architecture is the global controller, which takes the responsibility to generate control signals to guarantee the whole system to function correctly.

\section{EVALUTATION}
\subsection{FPGA}

\begin{table*}[t]
\small
  \caption{Comparison results on accuracy, performance, and energy efficiency of the proposed SWM-based FPGA designs and baselines.}
  \vspace{-0.1in}
  \label{table:1}
  \centering
  \begin{tabular}{|c|c|c|c|c|c|c|}
    \hline
    \textbf{DNN Name} & \textbf{Dataset} &\textbf{ Platform} & \textbf{Data Quantization} &\textbf{ Accuracy} &\textbf{ Performance} &\textbf{ Energy efficiency}\\
         &         &          &           &          & (kFPS)      & (kFPS/W) \\
    \hline
    Proposed MNIST 1 & MNIST & CyClone V & 12 bits& 92.9\% & $8.6\times 10^4$ & $1.57\times 10^5$\\
    Proposed MNIST 2 & MNIST & CyClone V & 12 bits& 95.6\% & $2.9\times 10^4$ & $5.2\times 10^4$\\
    Proposed MNIST 3 & MNIST & CyClone V & 12 bits& 99.0\% & $363$ & 659.5\\
    Proposed SVHN & SVHN & CyClone V & 12 bits& 96.2\% & $384.9$ & 699.7\\
    Proposed CIFAR-10 1 & CIFAR-10 & CyClone V & 12 bits& 80.3\% & $1383$ & 2514\\
    Proposed CIFAR-10 2 & CIFAR-10 & CyClone V & 12 bits& 94.75\% & $13.95$ & 25.4\\
    \hline
    TrueNorth (\cite{esser2015backpropagation}) & MNIST & TrueNorth & 2 bits & 99\%+ & 1.0 & 9.26 \\
    TrueNorth (\cite{esser2015backpropagation}) & MNIST & TrueNorth & 2 bits & 95\% & 1.0 & 250 \\
    TrueNorth (\cite{esser2016convolutional}) & SVHN & TrueNorth & 2 bits & 96.7\% & 2.53 & 9.85 \\
    TrueNorth (\cite{esser2016convolutional}) & CIFAR-10 & TrueNorth & 2 bits & 83.4\% & 1.25 & 6.11 \\
    \hline
    \hline
    \textbf{LSTM Name} &\textbf{ Dataset }&\textbf{ Platform} &\textbf{ Data Quantization} & \textbf{PER Degradation} &\textbf{ Performance} &\textbf{ Energy efficiency}\\
        \hline
     Proposed LSTM1 & TIMIT & ADM-7V3 & 16 bits & 1.23\% & 330.275 & 14.359 \\
     Proposed LSTM1 & TIMIT & KU060 & 16 bits & 1.23\% & 371.095 & - \\
     Proposed LSTM2 & TIMIT & ADM-7V3 & 16 bits & 0.32\% & 179.687 & 8.168 \\
     Proposed LSTM2 & TIMIT & KU060 & 16 bits & 0.32\% & 195.312 & - \\
    \hline
     ESE~\cite{han2017ese} & TIMIT & KU060 & 12 bits & 0.30\% & 17.544 & 0.428 \\

    \hline
  \end{tabular}
\end{table*}

 We implement the proposed framework on small to medium scale DNNs using the benchmarks of MNIST, SVHN, and CIFAR-10 on the low-power FPGA Intel (Altera) CyClone V 5CEA9.
And we implement the proposed method  LSTM on the platforms of Xilinx KU060 and Alpha Data's ADM-7V3. The ADM-7V3 board contains a Xilinx Virtex-7 (690t) FPGA and a 16GB DDR3 memory and the Xilinx KU060 platform contains a Xilinx XCKU060 FPGA and two 4GB DDR3 memory. We connect the ADM-7V3 to the host through PCI-e 3.0 X8 interface and the host machine used in the experiment is a sever configured with an Intel Core i7-4790 CPU. The proposed FPGA implementations of LSTMs are operating at 200MHz on both platforms.

We compare the accuracy, performance (kFPS), and energy efficiency (kFPS/W) of the proposed SWM-based FPGA implementation with the state-of-the-art IBM TrueNorth neurosynaptic processor (\cite{esser2015backpropagation})  for DCNNs, and the state-of-the-art ESE accelerator on the platform of Xilinx KU060~\cite{han2017ese} for LSTMs. We first demonstrate the results of three MNIST datasets targeting at different accuracies, one SVHN dataset, and two CIFAR-10 datasets targeting at different accuracies. The first two DNNs of MNIST datasets are multi-layer perceptron (MLP) models which can achieve the accuracy of 92.9\% and 95.6\%, respectively. The third DNN of MNIST dataset has a CNN structure similar to LeNet-5~\cite{lecun1995comparison}, which achieves 99.0\% accuracy. The first DNN of CIFAR-10 has a simple structure while the second DNN of CIFAR-10 adopts a wide ResNet model~\cite{he2016deep} which can achieve 94.75\% accuracy. The baseline system (IBM TrueNorth) has two different DNNs of MNIST datasets at two accuracy levels. Experimental results show that under the similar accuracy constraint, the gains of the SWM-based framework in performance and energy efficiency are at least 152X and 72X, respectively. For the LSTM implementation, we propose two structures: (i) the proposed LSTM1 adopts a block size of 16 (FFT16), which the relative PER degradation of the model is 1.23\%; (ii) the proposed LSTM2 uses a block size of 8 (FFT8), which the relative PER degradation of the model is 0.32\%. On the platform of KU060, we achieve 21X and 11X  performance speedup for the proposed LSTM1 and LSTM2 based compression techniques compared with ESE. On the platform of AMD-7v3, compared with ESE, we achieve 18.8X and 10.2X and performance enhancement and  33.5X and 19.1X energy efficiency gains using the proposed LSTM1 and LSTM2, respectively. Since the power consumption of SWM-based LSTM is only half of the ESE, the energy efficiency gain is higher than performance. Please note that the manufacturing process of XCKU060 FPGA is 20nm while the process of Virtex-7 is 28nm, which means the actual energy efficiency gain should be more than the report here.

\subsection{ASIC}
In this work, we implement an ASIC design of the SWM-based neural network for the image recognition task, and it is tested with the MNIST dataset. The implemented neural network has the original structure of $512 \times 512-512 \times 512 - 512 \times 64 - 64 \times 10$, and this network is transferred into an SWM-based structure. The FFT module implemented in this work is a 64-point FFT, that is, it takes a vector of 64 real value numbers as inputs and generates their frequency domain representations. Consequently, the weight matrices has the structure of $8 \times 8 \times 64 - 8 \times 8 \times 64 - 1 \times 8 \times 64 - 64 \times 10$, where $(m \times n \times s)$ represents the weight matrix has $m$ rows and $n$ columns, and each element is a vector containing $s$ weights ($s$ is 64 in this case). Our weight matrix transformation is not applied to the output layer, so the weights in this layer still keep the original structure of $64 \times 10$.

Our ASIC design is implemented with SMIC 40nm technology (including memories) and synthesized with Synopsys Design Compiler 2016. Table \ref{tbl:asic_performance} shows the hardware performance of our design. It can be observed from the table, the SWM-based neural network exhibits impressive advantages in terms of power ($0.14 W$), throughput ($1.14\times10^6 Images/s$), and energy efficiency ($8.08\times10^6 Images/J$), suggesting that this method is greatly suitable for applying DNNs onto mobile/IoT devices.

\begin{table}[]
\centering
\caption{Hardware Performance of SWM-Based Neural Network Implemented in ASIC}
\label{tbl:asic_performance}
\resizebox{2.2in}{!}{
\begin{tabular}{|c|c|}
\hline
\bf{Metrics}                    & \bf{Performance}  \\\hline
Clock Frequency (MHz)           & $200$             \\
Area ($mm^2$)                   & $1.3$             \\
Power ($W$)                     & $0.14$           \\
Throughput ($Images/s$)         & $1.14\times10^6$  \\
Energy Efficiency ($Images/J$)  & $8.08\times10^6$  \\

\hline
    \end{tabular}
}
\end{table}

\section{Conclusion}
In this work, we propose and evaluate the SWM-based compression technique on both FPGA and ASIC implementations. The SWM-based framework adopts the general block-circulant matrices to achieve a fine-grained tradeoff of accuracy and compression ratio and it works for both FC and CONV layers and contains a mathematically rigorous proof.
For FPGA implementations, we achieve at least 152X and 72X improvement in performance and energy efficiency, respectively using SWM-based framework, compared with the baseline of IBM TrueNorth processor under same accuracy constraints using the data set of MNIST, SVHN, and CIFAR-10. For the LSTM network, the proposed SWM-based LSTM can achieve up to 21X enhancement in performance and 33.5X gains in energy efficiency compared with ESE, respectively. For ASIC implementations,  the proposed SWM-based design exhibits impressive advantages in terms of power, throughput, and energy efficiency. Experimental results indicate that this method is greatly suitable for applying DNNs onto both FPGAs and mobile/IoT devices.

\section{Acknowledgement}
 This work is funded by the National Science Foundation
Awards CNS-1650469, CCF-1733701, CNS-1704662,
CCF-1657333, CNS-1739748, and CCF-1733834.

\bibliographystyle{ACM-Reference-Format}
\bibliography{sample-bibliography}

\end{document}